\documentclass[%
 reprint,
superscriptaddress,
 amsmath,amssymb,
 aps,
 prl,
]{revtex4-1}

\usepackage[final]{changes}
\usepackage{upgreek}
\usepackage{graphicx}
\usepackage{dcolumn}
\usepackage{bm}
\usepackage{setspace}
\usepackage{amsmath}
\usepackage[english]{babel}
\usepackage{lipsum}
\usepackage{float} 

\addto\captionsenglish{}

\newcommand{\nv}{NV$^-$}
\newcommand{\nvs}{NV$^-$s~}

\begin{document}


\preprint{APS/123-QED}

\title{High-sensitivity, spin-based electrometry with an \\ensemble of nitrogen-vacancy centers in diamond}


\author{Edward H. Chen}
\affiliation{Dept. of Electrical Engineering and Computer Science, Massachusetts Institute of Technology (MIT)}
\affiliation{MIT Lincoln Laboratory, Lexington, MA 02420, USA.}
\affiliation{Present Address: HRL Laboratories, LLC. Malibu, CA 90265, USA.}
\author{Hannah A. Clevenson}
\affiliation{Dept. of Electrical Engineering and Computer Science, Massachusetts Institute of Technology (MIT)}
\affiliation{MIT Lincoln Laboratory, Lexington, MA 02420, USA.}

\author{Kerry A. Johnson}
\affiliation{MIT Lincoln Laboratory, Lexington, MA 02420, USA.}

\author{Linh M. Pham}
\affiliation{MIT Lincoln Laboratory, Lexington, MA 02420, USA.}

\author{\\Dirk R. Englund}
\affiliation{Dept. of Electrical Engineering and Computer Science, Massachusetts Institute of Technology (MIT)}

\author{Philip R. Hemmer}
\affiliation{Dept. of Electrical and Computer Engineering, Texas A\&M University (TAMU)}

\author{Danielle A. Braje}
\email{braje@ll.mit.edu}
\affiliation{MIT Lincoln Laboratory, Lexington, MA 02420, USA.}

\date{\today}
\begin{abstract}

We demonstrate a spin-based, all-dielectric electrometer based on an ensemble of nitrogen-vacancy (\nv) defects in diamond. An applied electric field causes energy level shifts symmetrically away from the \nv's degenerate triplet states via the Stark effect; this symmetry provides immunity to temperature fluctuations allowing for shot-noise-limited detection. Using an ensemble of \nv s, we demonstrate shot-noise limited sensitivities approaching 1~V/cm/$\sqrt{\text{Hz}}$ under ambient conditions, at low frequencies ($<$10~Hz), and over a large dynamic range (20 dB). A theoretical model for the ensemble of \nv s fits well with measurements of the ground-state electric susceptibility parameter, $\langle k_\perp\rangle$. Implications of spin-based, dielectric sensors for micron-scale electric-field sensing are discussed.



\end{abstract}

\maketitle

\section{Introduction}\vspace{-5pt}
\label{sec:intro}
The detection of weak electric signals in low-frequency regimes is important for areas of research such as particle physics~\cite{sharma_diamond-based_2016}, atmospheric sciences~\cite{merceret_magnitude_2008,simoes_review_2012,mezuman_spatial_2014}, and neuroscience~\cite{barry_optical_2016}. Commonly available ambient electrometers that rely on electrostatic induction, like field mills~\cite{fort_design_2011,riehl2003electrostatic} and dipole antennas~\cite{harrington_effect_1960}, are physically limited in size to several tens of centimeters by the wavelength of the electric field of interest. This hinders miniaturization at frequencies below several Hertz~\cite{chu_physical_1948,harrington_effect_1960,chubb_limitations_2015,barr_elf_2000}. Fully dielectric sensors allow sensing of electric fields without fundamental constraints on the size of the sensor and do not distort the incident field~\cite{savchenkov_photonic_2014,toney_lithium_2015}. Ongoing efforts to develop compact electrometers include use of the electro-optic effect within solid-state crystals~\cite{vohra_fiber-optic_1991}, single electron transistors~\cite{lee2008room,vincent2004theory,neumann2013graphene}, and the energy shifts induced by electric fields of atom-based sensors such as trapped ions~\cite{bruzewicz_measurement_2015} or Rydberg atoms~\cite{osterwalder_using_1999,sedlacek_microwave_2012}. Recently, optically-addressable electron spins in solid-state materials have played a central role in the development of quantum sensing~\cite{awschalom_quantum_2013,degen2016quantum}. Compared with atom-based approaches that require vacuum systems, these systems allow for a higher density of spins with a reduced experimental footprint, along with other promising properties such as long room-temperature coherence times and optical accessibility for spin initialization and readout~\cite{acosta2013}.



\begin{figure}[!b]
\vspace{-30pt}
		\includegraphics[width=85mm]{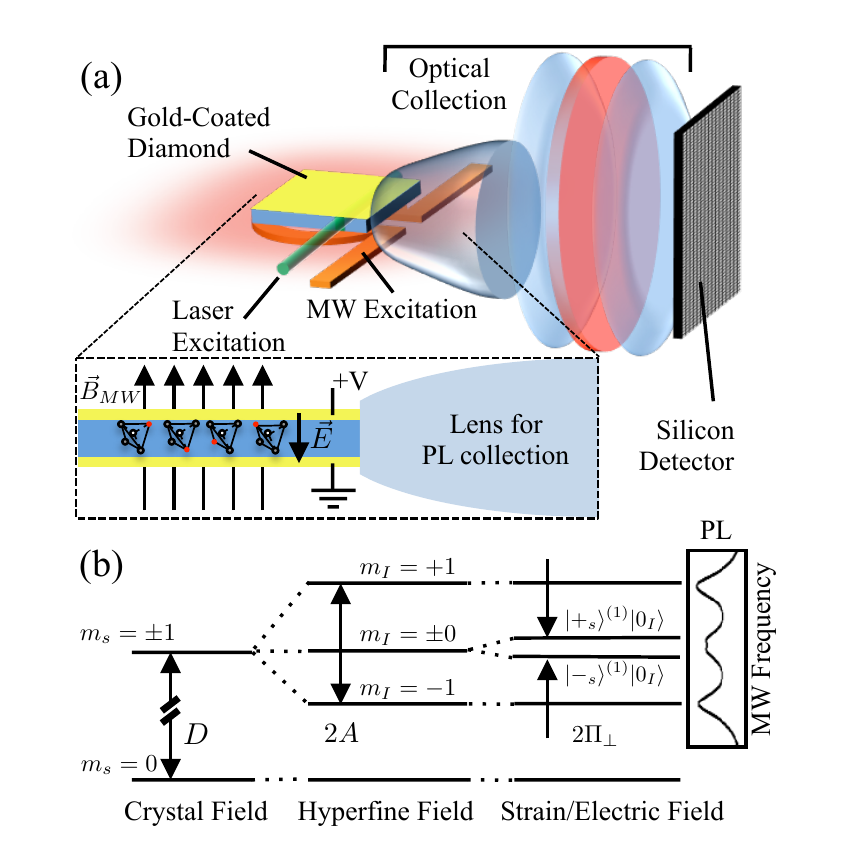}
	\caption{ 
	(a) Diamond electrometry setup. For applying electric fields across the ensemble of \nv s, gold electrodes were evaporated on both faces of the diamond plate (3 x 3 x 0.32~$\text{mm}^3$). A collimated laser beam ($\sim$ 200~$\upmu$m diameter) was used to excite a single pass of \nvs~input from the edge of the diamond plate, and microwave (MW) excitation was delivered to the ensemble of \nvs~by a stripline in an `$\Omega$' shape patterned on a printed circuit board. \textit{Inset} A cross-section of the experiment depicting four of the eight total \nv orientations within an ensemble of \nvs used for detecting electric fields.
	(b) Generalized diagram depicting how crystal ($D$), hyperfine ($A$), strain and the magnitude of transverse electric fields ($\Pi_\perp = \sqrt{\Pi_x^2+\Pi_y^2}$) affect energy splitting in both the ground- and excited-state spin configurations of the \nv. The spin labels ($m_s$ and $m_I$) indicate the quantum numbers of the electronic and hyperfine states, and the two eigenstates that are sensitive to electric fields are given by:~$| + \rangle_s^{(1)} | 0 \rangle_I$ ($m_I = +0$) and $| - \rangle_s^{(1)} | 0 \rangle_I$ ($m_I = -0$). The inset to the right shows the ground-state ODMR spectra of an ensemble of \nv s.
	}
	\vspace{-30pt}
\label{fig:Fig1_overview}
\end{figure}
Among spin-based sensors, there has been significant progress in using ensembles of spins in diamond for sensing magnetic fields~\cite{clevenson_broadband_2015,fang_high-sensitivity_2013,barry_optical_2016}, while work in diamond-based electrometry has primarily focused on the use of single spins~\cite{dolde_electric-field_2011,doherty_measuring_2014}. Here, we experimentally demonstrate a spin-based, solid-state electrometer that is sensitive to the electric field induced Stark shift on an ensemble of negatively-charged nitrogen-vacancy (\nv) color centers while being robust, to first order, to temperature fluctuations. Our diamond-based electrometer operates at shot-noise limited sensitivities of $\approx1~\text{V/cm}/\sqrt{\text{Hz}}$ under ambient conditions at extremely low frequencies (0.05-10~Hz) without repetitive readout and dynamic decoupling control necessary for electrometry with single \nv s~\cite{dolde_electric-field_2011}. By utilizing a high degree of symmetry to overcome the inhomogeneous strain and non-collinear crystallographic orientations within an ensemble of \nv s, this work brings diamond-based electrometry into a regime where it has a competitive sensitivity with a clear path towards miniaturization.

\nv~centers are sensitive to electric fields in both their optical ground~\cite{dolde_electric-field_2011} and excited triplet states~\cite{tamarat_stark_2006}. Previously, electric field sensing with a single \nv~was demonstrated with sensitivity down to 202~V/cm/$\sqrt{\text{Hz}}$~(891~$\text{V/cm}/\sqrt{\text{Hz}}$) at a frequency of $\sim10$~kHz (DC) under precisely applied magnetic fields, but the need for repetitive readout and dynamic decoupling pulse control limited that technique to frequencies in excess of 10~kHz due to the \nv 's decoherence rate  (1/T$_2$ where T$_2\sim~0.1$~ms). The device demonstrated here uses an ensemble of \nv~centers in an otherwise similarly sized diamond. It is not only possible to achieve higher sensitivities (albeit over a larger volume), but it also allows for a measurement of the noise spectral density (NSD) due to low-frequency electric field fluctuations irrespective of temperature fluctuations. Furthermore, the introduced method allows for highly accurate measurement of the transverse electric susceptibility parameter, $\langle k_\perp\rangle$, of the \nv's ground state. By using this measurement modality, we expect that a diamond that is densely populated with \nvs would yield a \textit{projected} shot-noise-limited electric field sensitivity approaching $6\times10^{-3}~\text{V/cm}/\sqrt{\text{Hz}}$~\footnote{See Supplemental Material at [URL will be inserted by publisher] for this projected sensitivity using values from recent magnetometry experiments~\cite{barry_optical_2016}.}, making \nv-based electrometers comparable with currently existing, room-temperature, solid-state electrometers~\cite{savchenkov_photonic_2014,toney_lithium_2015}.

\section{Theory of \nv~electrometry}
\label{sec:theory}
\vspace{-10pt}

The physical mechanism of the \nv's~sensitivity to electric fields originates from its optical excited state configuration, which is a highly electric field sensitive molecular doublet ($^3E$). Stark shifts of the excited state cannot be measured optically under ambient conditions due to phonon-induced mixing~\cite{plakhotnik_all-optical_2014}. However within each orbital of the molecular doublet, the electric field induced splitting of the $m_I = 0$ hyperfine manifold can be detected by optically detectable magnetic resonance (ODMR). Additionally, the $^3E$ excited-state orbital overlaps sufficiently with the ground-state molecular orbital ($^3A_2$) to also impart electric field sensitivity on the ground state spin configuration of the \nv~\cite{doherty_measuring_2014}.

\begin{figure}[!hb]
	\includegraphics[width=85mm]{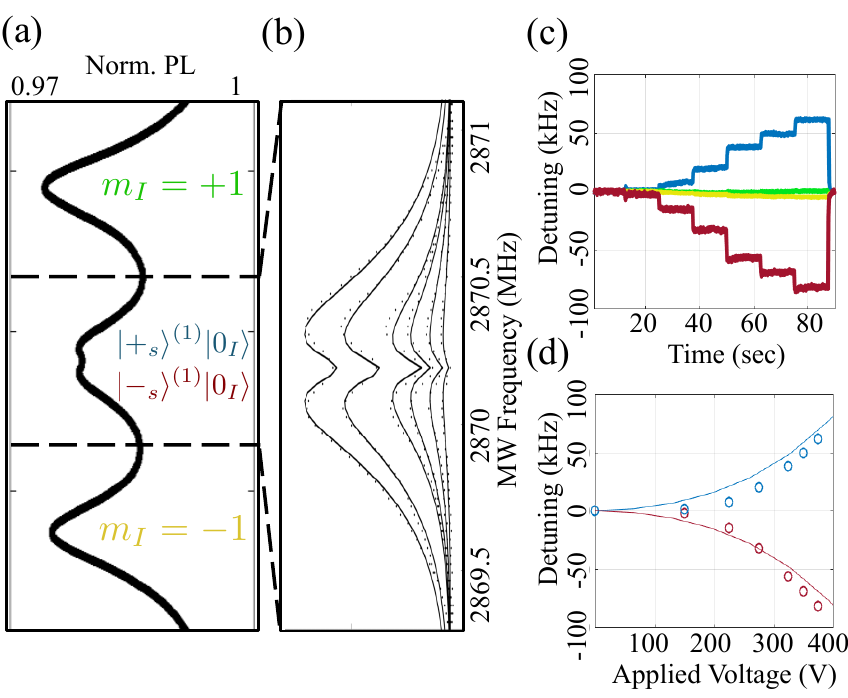}
	\caption{ 
	(a) ODMR spectrum with correspondingly colored labels to indicate the detuning of transitions with step-wise increasing applied voltages. 
	(b) Experimentally measured ODMR (points) at six different MW driving amplitudes overlaid with their respective numerical fits (black) using a model that accounts for an isotropic distribution of strain fields within an ensemble of \nv s. The two transitions correspond to the $| - \rangle_s^{(1)} | 0 \rangle_I$ (red, bottom) and $| + \rangle_s^{(1)} | 0 \rangle_I$ (blue, top) eigenstates of the \nv~triplet ground state. 
	(c) Ground-state shifts due to incremental, step-wise electric fields applied to an ensemble of \nv s at zero magnetic field. By comparing the step-wise detuning shifts of the electric field transitions with the applied voltages, it is possible to accurately deduce the ensemble average value of $\langle k_\perp \rangle = 7.0 \pm 1.1~\text{Hz}/\text{V/cm}$ at a bias field of 225~Volts.
	(d) Data from (c) represented as voltages and overlaid with numerical results~(See Eqn.~S2).
	}
\label{fig:Fig2_odmr}
\end{figure}

The Hamiltonian describing both the triplet ground and excited states of the \nv~share the following form~\cite{doherty_nitrogen-vacancy_2013}:

\begin{equation}
\hat{H}_{NV}/h = (D+d_\parallel E_\parallel)\hat{S_z}^2 
+ \gamma_B \vec{B}\cdot \mathbf{{g}} \cdot\vec{\hat{S}}
+ \vec{\hat{S}}\cdot \mathbf{A} \cdot\vec{\hat{I}}, \\
\label{eqn:nv_hamiltonian}
\end{equation}
 
\noindent where $D$ is the crystal field splitting (Hz), $\gamma_B$ is the gyromagnetic ratio ($\text{Hz}/\text{G}$), $d_\parallel$ is the axial electric field dipole moment ($\text{Hz}/\text{V/cm}$), $E_\parallel$ is the axial electric field (V/cm), $\vec{B}$ is the magnetic field vector, $\mathbf{g}$ is the g-factor tensor, $\vec{\hat{S}}$ is the vector of electronic spin-1 Pauli operators, $\mathbf{A}$ is the hyperfine tensor, and $\vec{\hat{I}}$ is the vector of nuclear spin-1 Pauli operators. Because an electric field's effect on the \nv~spin is significantly smaller than the crystal field splitting ($D$), the transverse electric field dependence can be considered as a perturbation to the Hamiltonian:

\begin{equation}
\hat{V}/h = d_\perp[\Pi_x(\hat{S_x}\hat{S_y}+\hat{S_y}\hat{S_x})+\Pi_y(\hat{S_x^2}-\hat{S_y^2})],
\label{eqn:nv_perturbation}
\end{equation}

\noindent where $d_\perp$ is the ground-state's transverse electric field dipole moment~\footnote{See~\cite{doherty_measuring_2014} for the relationship between electric-dipole moment, $d_\perp$ and the electric susceptibility, $k_\perp$.}, $\Pi_x$ and $\Pi_y$ are the cartesian components of the combined strain and electric fields~\footnote{See Fig.~7 for the differences and similarities in using the ground and excited states of the \nv for electrometry}, and $\hat{S_i}$ (for $i=x,y,z$) are the spin-1 Pauli operators of the electronic spin. After diagonalizing Eqn.~\eqref{eqn:nv_hamiltonian} and using Eqn.~\eqref{eqn:nv_perturbation} as the perturbation, there is a closed-form equation which describes the effect of electric and magnetic fields on the \nv~(See Eqn.~S1). The following equation accurately describes how the eigenfrequencies of a single \nv change under an applied electric field ($\vec{E}$) alongside no magnetic field ($\vec{B}=0$). Furthermore, the expression quantitatively matches the transition shifts due to a symmetric application of electric fields on all eight classes of defects within an ensemble of \nv~centers:

\begin{equation} 
\label{eqn:analytical}
f_{\pm}(\vec{B}=0,\vec{E}) = D + k_\parallel E_\parallel \pm k_\bot E_{\bot}
\end{equation}
where $D$ is the crystal field splitting with a temperature dependence of $\approx$77~kHz/Kelvin~\cite{chen2011temperature}, $k_\parallel$ and $k_\bot$ are the electric susceptibility parameters (in units of $\text{Hz}/\text{(V/cm)}$), and $E_\parallel$ and $E_{\bot}$ are the electric field amplitudes (in units of V/cm) parallel and perpendicular, respectively, to the \nv~symmetry axis. 

The shot-noise sensitivity to the transverse electric field (in $\text{V/cm}/\sqrt{\text{Hz}}$) limits using an ensemble of $M$ \nvs~is given by the following:
\begin{equation}
\eta_{E_\perp} \approx \frac{1}{k_{\perp}} \frac{1}{C \sqrt{M \Gamma}}\frac{1}{T_2^*} 
\label{eqn:shot_noise_lim} 
\end{equation}
where $C$ is the contrast of the ensemble ODMR spectra, $\Gamma$ is the total photon collection rate per \nv, and $T_2^*$ is the inhomogeneous \nv~coherence time. Using our experimentally measured values, we arrive at a shot-noise limit approaching $\eta_{E_\perp} \approx 1.0~\text{V/cm}/\sqrt{\text{Hz}}$ for the \nv~ground state~\footnote{$k_\perp^{225~\text{V Bias}} = 7~\text{Hz}/\text{V/cm}, C\approx0.02, 1/T_2^* \approx 4\times10^5 \text{~Hz}, M \approx 5\times10^9, \gamma \approx 100~\text{photons per second}$ }. This expression shows that the sensitivity limit depends on the density of \nvs in the sample, the coherence properties of the \nv s, and the efficiency of photon collection.

\section{Experimental Results}
The diamond measured in this work is 3.0~x~3.0~x~0.32~$\text{mm}^3$ in size and contains an \nv~density of $\sim$1~ppb produced during the chemical vapor growth process. The two square faces on which the electrodes were evaporated have $(100)$ crystallographic orientations~(See Fig.~1), and thus the applied electric field produces an equal projection onto all eight orientations of NVs within the ensemble. We measured ODMR of the ensemble of \nvs~using continuous-wave (CW) laser and microwave excitation from the side and bottom, respectively. 

To account for the distribution of strain magnitudes and angles within the ensemble of \nv s, we use the (2,1) Gamma probability distribution as an ansatz for the magnitude distribution, and assume a uniform and isotropic angular distribution~\footnote{See Supplemental Material at [URL will be inserted by publisher] for an analysis considering the isotropic distribution of strain with an ensemble of \nv s.}. This model accounting for the isotropic distribution of strain accurately fits the experimental data (See Fig.~2b). We also simulated the expected electric field induced shift on the ensemble average of \nvs~in Fig.~2c using an estimated distribution of strain. The simulated results match well with the step-wise increase in electric field; this agreement validated the use of this method for accurately scaling the measured shift in frequency to the reported noise floor of the noise spectral density~(See Figures~\ref{fig:Fig2_odmr}c,d).

To detect a shift in \nv~transition frequencies of the \nv ensemble due to an external electric field, the magnetic field along the \nv~axes must be significantly weaker than the internal electric or strain fields. The maximum electric field sensitivity is achieved at zero magnetic field, but at the expense of vector sensitivity~\cite{dolde_electric-field_2011}. Additionally, the shot-noise limited sensitivity given by Eqn.~\ref{eqn:shot_noise_lim} can be optimized by controlling the laser and microwave excitation powers. 

\begin{figure}[H]
	\centering
		\includegraphics[width=85mm]{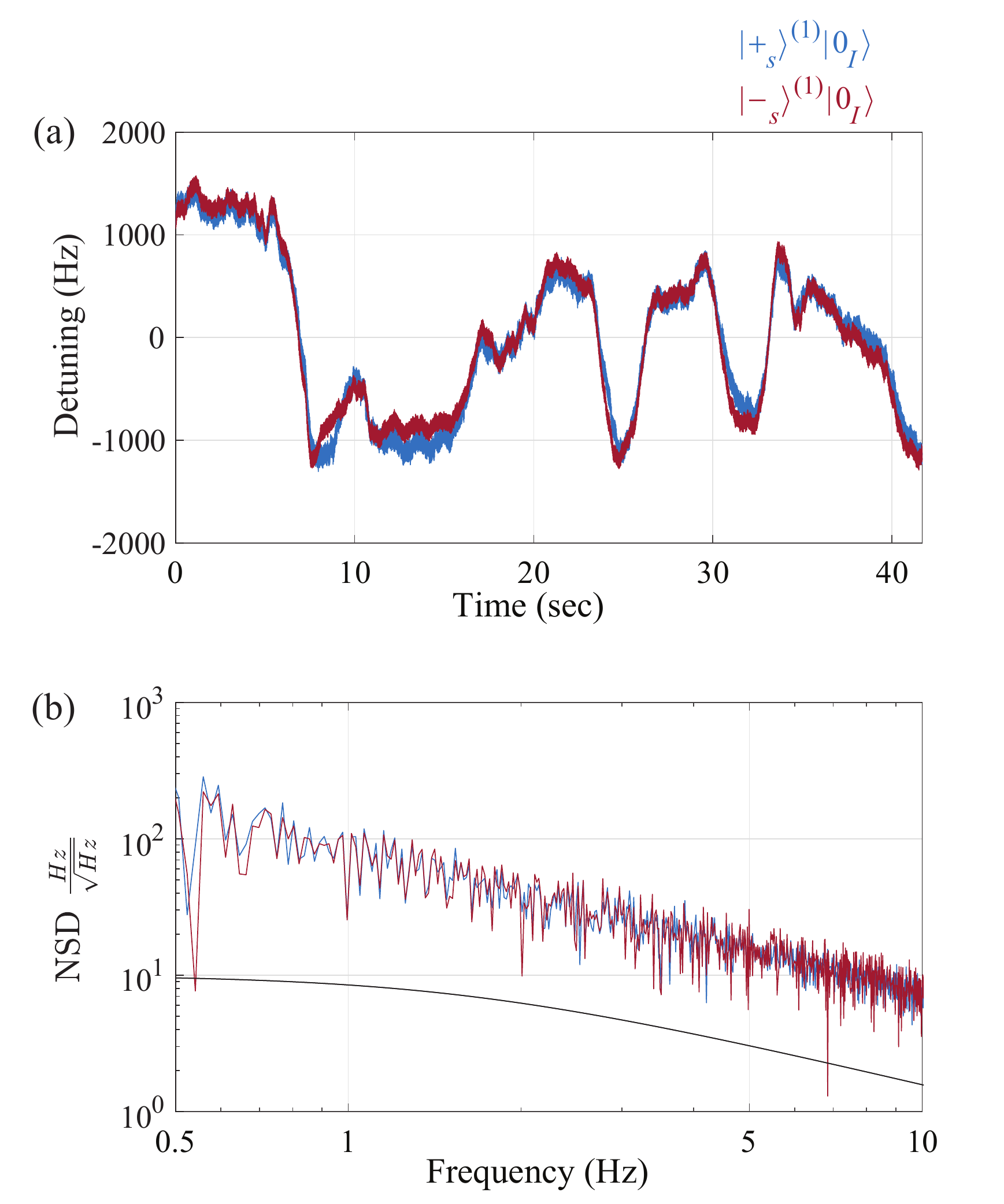}
	\caption{ Measurements taken at 1.8~W laser excitation with a high-stability bias voltage of 225~Volt.
	(a) Time trace of both lock-in-amplifier channels monitoring frequency detuning of ensemble states $| + \rangle_s^{(1)} | 0 \rangle_I$ (blue) and $| - \rangle_s^{(1)} | 0 \rangle_I$ (red) in units of transition frequency noise per $\sqrt{\text{Hz}}$. 
	(b) Noise spectral density (NSD) of both channels. The noise floor of the red (blue) channel is calculated to be equivalent to $12.6\pm6.4$~($13.4\pm7.4$)~$~\text{V/cm}/\sqrt{\text{Hz}}$, assuming the noise is entirely attributed to electric field fluctuations. The electric field sensitivity estimated by the shot-noise limit is given by the black line ($1.2\pm0.1~\text{V/cm}/\sqrt{\text{Hz}}$).
	}
\label{fig:Fig3_2chan_lia}
\end{figure}

We measured two electric and strain sensitive transition frequencies (denoted as $m_I = \pm0$) simultaneously at a rate inversely proportional to the time constant of the home-built lock-in instrumentation~\footnote{See Supplemental Material at [URL will be inserted by publisher] for a description of the technical details of the setup.}. Although only the two transitions are monitored, the shift in frequencies correspond to an average shift due to the entire ensemble. The inhomogeneous strain typically found in the ensemble is indistinguishable from an inhomogeneous distribution of electric fields. Using a bias electric field beyond the average strain of the ensemble of \nv~centers, the shift of the $m_I = \pm0$~(See Fig.~1 for notation) transitions become linearly sensitive to electric fields, while the transitions, $m_I = \pm1$, remain relatively insensitive to electric fields due to the quadrupole field of the host nuclear $^{14}N$ spin~(See Fig.2d). 

Figure 3 presents the resulting sensitivity measurements. A maximum electric field sensitivity of the ensemble of \nvs was achieved with an incident laser power of 1.8~W. However, the high input laser ($\sim$30~$\upmu$W/$\upmu \text{m}^2$) powers required to saturate the photoluminescence from the \nv s contributes to greater temperature fluctuations in the diamond. In a simultaneous time trace of the $m_I = \pm0$ transitions, there are significant correlated shifts due to the temperature fluctuations (See Fig.~\ref{fig:Fig3_2chan_lia}a). The noise-spectral densities (NSD) of the two time traces indicate $1/f$-type noise, which is consistent with the source of the noise being due to temperature fluctuations. The noise floors of both channels are more than a factor of 10$\times$ greater than the shot-noise limit (See Fig.~3b).

The temperature fluctuations are separated from the electric field fluctuations using the temperature-dependent, correlated shifts of the $D$ parameter (See Eqn.~\ref{eqn:analytical}). The sum of the time traces corresponds to the temperature fluctuations while the difference of the time traces corresponds to the electric field fluctuations. The NSD of the resulting sum and differences shows temperature fluctuations of $2.4\pm1.2~\text{mK}/\sqrt{\text{Hz}}$ and electric field fluctuations of $1.6\pm1.2~\text{V/cm}/\sqrt{\text{Hz}}$, respectively~(See Fig.~4). Thus, our method shows a shot-noise-limited electric-field sensitivity that is approximately 8$\times$ improved over a measurement without deconvolution with temperature fluctuations.

\begin{figure}[!hb]
	\centering
		\includegraphics[width=85mm]{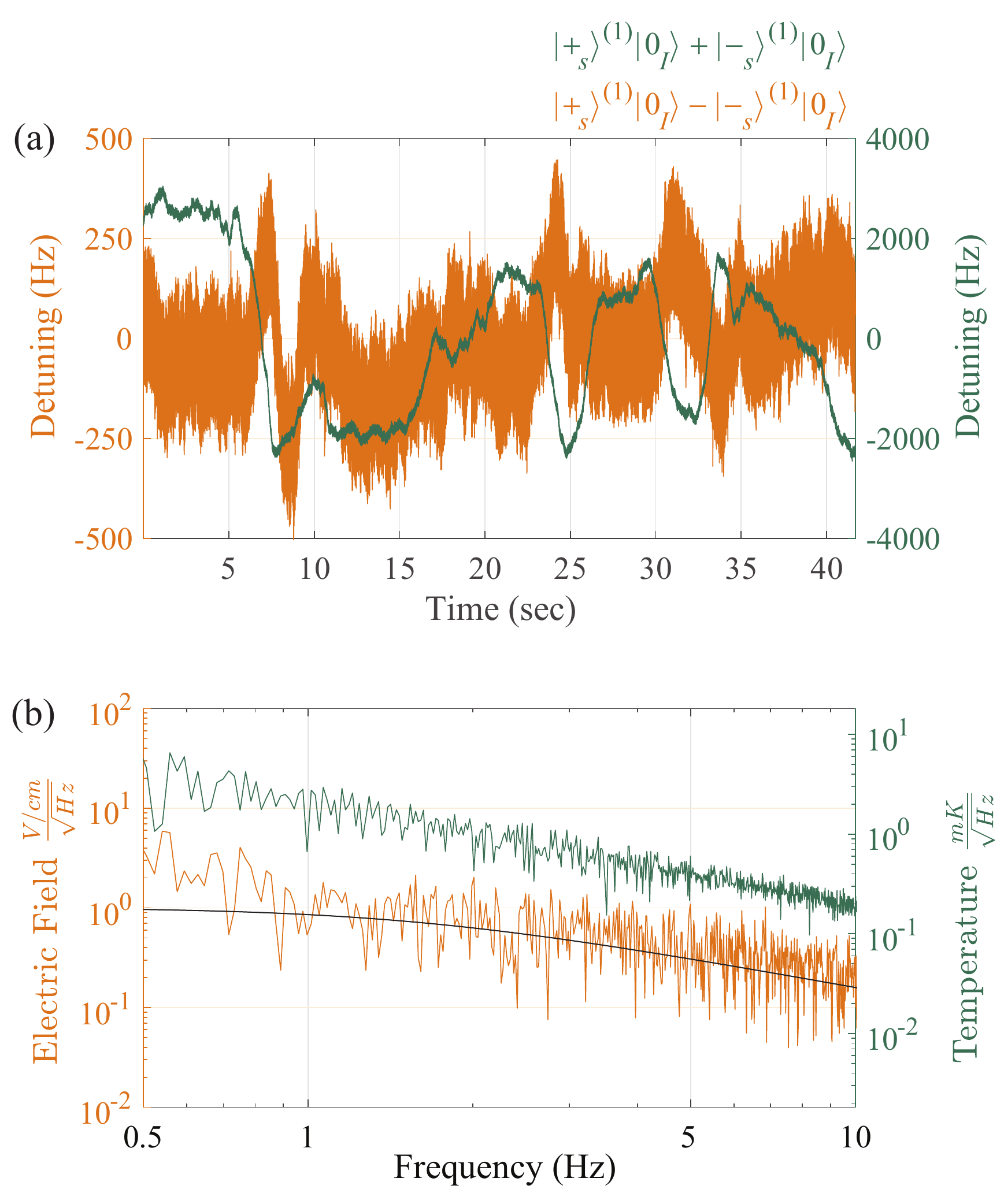}
	\caption{ Identical experimental measurements as in Fig.~3 except the analysis takes advantage of the experimental methodology for deconvolving fluctuations from temperature and electric fields.
	(a) Time trace of the difference (orange, electric field) and sum (green, temperature) of the lock-in-amplifier channels from Fig.~3.	
	(b) NSD on the time-trace difference (sum) of the two channels, which corresponds to a sensitivity of $1.6\pm1.2$~$~\text{V/cm}/\sqrt{\text{Hz}}$~($2.4\pm1.2~\text{mK}/\sqrt{\text{Hz}}$) due the transverse electric field (temperature) fluctuations. The sum of the two correlated channels yields a shot-noise sensitivity limit of $0.9\pm0.1~\text{V/cm}/\sqrt{\text{Hz}}$ (black line), which is $\sqrt{2}$ times lower than that of the individual channels as seen in Fig.~3b.
	}
\label{fig:Fig4_2chan_lia}
\end{figure}

\section{Discussion}
\vspace{-10pt}
We have demonstrated sensing of electric fields with an ensemble of \nvs below 1~Hz with sensitivities approaching $1~\text{V/cm}/\sqrt{\text{Hz}}$. In spite of large temperature variations, inhomogeneous distribution of strain and non-collinear orientations, our measurement technique allows for accurate measurements of the ensemble strain distribution and the ensemble average of the transverse electric susceptibility, $k_\perp$; both of which are needed to accurately measure low-frequency electric fields. 

\nv-based sensing lends itself to imaging electric fields at or below the optical diffraction limit~\cite{chen_wide-field_2013,pfender_single-spin_2014,hsiao_fluorescent_2016}. We anticipate that the use of low-strain nanodiamonds with our demonstrated zero-magnetic field regime would enable simultaneous monitoring of both temperature~\cite{kucsko_nanometre-scale_2013-1} and electric fields~\cite{dolde_electric-field_2011}. To the best of our knowledge, nanodiamonds with low-strain ($<200$~kHz) are not yet available despite the tremendous progress in improving the electronic coherence within such nano-scale structures~\cite{knowles_observing_2014,trusheim_scalable_2014}. Such low-strain nanodiamonds with high densities of \nvs~would be beneficial for \textit{in vitro} biological studies~\cite{rehor_nanodiamonds:_2016,karaveli_modulation_2016} and microelectronic diagonistics~\cite{nowodzinski_nitrogen-vacancy_2015}. Finally, due to the many combinations of host materials and defects, there is significant potential in discovering defects within two- and three-dimensional materials that would further improve upon existing electronic spin-based electrometers~\cite{freysoldt_first-principles_2014,tran_quantum_2016}.

In this work, we have demonstrated a factor of more than 200x improvement over previous demonstrations using a single \nv. The sensitivity may be further improved by using a diamond with 1000x higher densities of \nv s~\cite{barry_optical_2016}, improving the photon collection efficiency by another 10-100 times by patterning the diamond surface to overcome the confinement due to total internal reflection~\cite{schell_laser-written_2014,li_efficient_2015}, and implementing pulsed control techniques to avoid power-broadening of the transitions~\cite{hodges_timekeeping_2013,toyli_fluorescence_2013,neumann_high-precision_2013,jamonneau_competition_2015}. Such readily accessible material and setup improvements could improve the shot-noise limited electric field sensitivity to ~$6\times10^{-3}~\text{V/cm}/\sqrt{\text{Hz}}$. Additional coherent control on either the surrounding electron~\cite{cappellaro_spin-bath_2012,bonato_optimized_2015} or nuclear spins~\cite{hirose_coherent_2016,unden_quantum_2016,clevensonPRA2016} in diamond would further improve the sensitivity by reducing the broadening of the transition line width. MW field inhomogeneities that are typically more problematic for pulsed techniques would benefit from recently proposed methodologies for generating robust pulse sequences~\cite{farfurnik2015optimizing}. Other promising directions for spin-based sensing involve all-optical techniques in diamond for electrometry~\cite{wickenbrock_microwave-free_2016}. 

\acknowledgements{Acknowledgements: The authors would like to thank Paola Cappellaro, John Barry, John Cortese, Colin Bruzewicz, Florian Dolde, Carson Teale, Christopher McNally, Christopher Foy, Xiao Wang, Peter Murphy, Sinan Karaveli, Jeremy Sage, and Jonathon Sedlack for fruitful discussions. E.H.C. and H.C. were supported by NASA's Office of Chief Technologist on the Space Technology Research Fellowship. D.E. acknowledges support by the Air Force Office of Scientific Research PECASE, supervised by Dr. Gernot Pomrenke.

This material is based upon work supported by the Assistant Secretary of Defense for Research and Engineering under Air Force Contract No. FA8721-05-C-0002 and/or FA8702-15-D-0001. Any opinions, findings, conclusions or recommendations expressed in this material are those of the author(s) and do not necessarily reflect the views of the Assistant Secretary of Defense for Research and Engineering.}

\vspace{-10pt}
\bibliography{NVElectrometry}
\newpage
\appendix

\section{Supplementary Information}

\subsection{Addressing an ensemble of \nvs}

\begin{figure}[H]
	\centering
		\includegraphics[width=90mm]{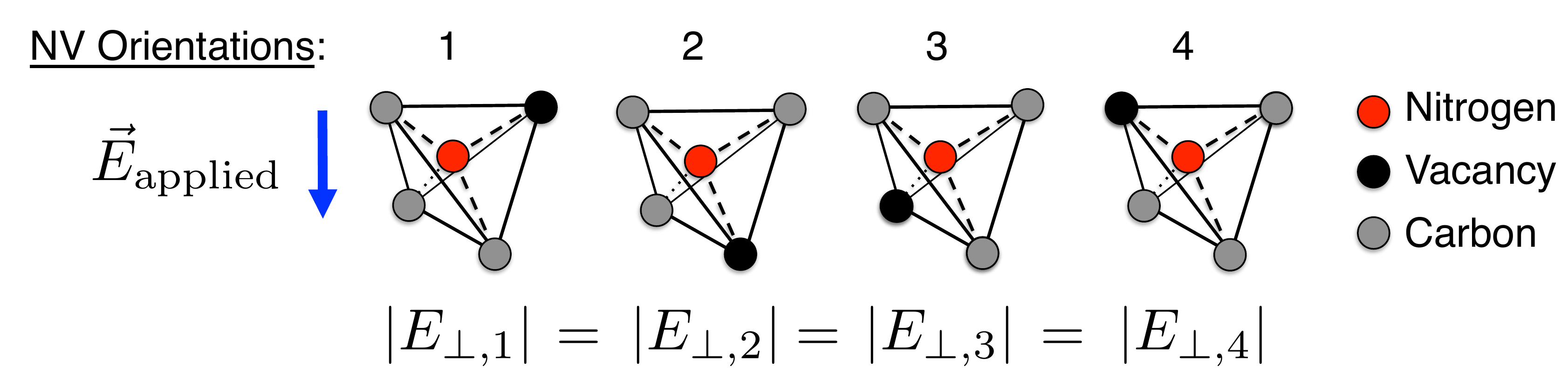}
	\caption{Applied electric field with respect to the eight orientations of NVs in the ensemble.
	}
\label{fig:4orientations}
\end{figure}

A diagram showing four of the eight possible \nv~orientations found with a diamond containing an ensemble of \nvs~(See Fig.~\ref{fig:4orientations}).

\subsection{Sensitivity approaching $\frac{\text{mV}}{\sqrt{\text{Hz}}}$}
Using Eqn.~4 in the main text, we expect a shot-noise-limited sensitivity approaching \\$6\times10^{-3} \frac{\text{V/cm}}{\sqrt{\text{Hz}}}$ using photocurrent values of 10~mW ($n \gamma = $~$62\times 10^{15}$ eV/sec $\times$ 1~photon/1.9eV = $3 \times 10^{16}$ photon/sec) as typically seen with ensemble \nv~measurements for magnetometry experiments~[5], a transverse electric susceptibility of $k_\perp = 17 \frac{Hz}{\text{V/cm}}$, line width ($\Delta f$) of 1~MHz, and contrast($C$) of 0.05.

\subsection{Full energy transition expression}
\label{supp:full_mw_expression}
Using second-order, degenerate perturbation theory, we derive the microwave transition frequencies between the eigenstates split by transverse electric fields:

\begin{multline} 
\label{eqn:analytical_supp}
\omega_{\pm}(\vec{E},\vec{B}) = D + k_{\parallel} E_{\parallel} + 3 \frac{(\gamma_B B_{\bot})^2}{2D} \\ 
\pm \sqrt{B_{\parallel}^2+E_{\bot}^2-\frac{1}{2}\sqrt{B_{\parallel}^2+E_{\bot}^2}\frac{B_{\bot}^2}{2D} sin(\alpha)cos(\beta) + (\frac{B_{\bot}^2}{2D})^2}
\end{multline}
where $\text{tan} (\alpha) = E_{\bot}/B_{\parallel}$, $\beta = 2\phi_B + \phi_E$, $\text{tan} (\phi_B) = B_y/B_x$, $\text{tan} (\phi_E) = E_y/E_x$, $B_{\perp} \equiv \sqrt{B_x^2 + B_y^2}$, and $E_{\perp} \equiv \sqrt{E_x^2 + E_y^2}$.

The equation which describes the ensemble ODMR spectrum is given by:
\begin{widetext}
\begin{equation} 
\label{eqn:odmr_integral}
I_\pm(f) =
1 - \frac{1}{24 \pi }\int _0^{2 \pi }\int _0^{\pi }\int _0^{\infty }\frac{C_o~\textbf{P}(x)}{4 \left(\frac{f-(D_o \pm k_\bot E_o x \sin (\theta ))}{\Delta f_o}\right)^2+1} x^2 \sin (\theta ) dxd\theta d\phi
\end{equation}
\end{widetext}

where $f$ is the frequency of the applied MW field, $C_o$ is the ensemble average of the ODMR contrast, $D_o$ is the ensemble average of the crystal field, $\textbf{P}(x) = x e^{-x}$ is the (2,1) Gamma probability distribution of the strain magnitude, $E_o$ is the ensemble average of the strain magnitude, $\Delta f_o$ is the full-width half-maximum of single-NV line-widths, $\theta$ denotes the strain vector's altitude angle away from the \nv~symmetry axis, and $\phi$ is the strain vector's azimuthal angle. 

\subsection{Zero-ing of magnetic field using gradient descent}
It is possible to zero the magnetic field using gradient descent because the overlap of the $m_I = \pm1$ transitions of all eight orientations of \nvs has a contrast that varies smoothly with respect to applied small magnetic fields. By taking local gradients of the contrast at each magnetic field setting ($B_x$, $B_y$ and $B_{\parallel}$) followed by successively smaller step sizes, we find the setting of $\vec{B}$ that achieves the globally maximum ODMR contrast and hence a zero magnetic field.

\subsection{Digital Lock-In Amplifier Implementation}
Using an FPGA high speed DAC, our system contains both the waveform generation and lock-in detection to perform readout of the optical signals from the diamond. The MW waveform sent to the diamond is generated digitally in the FPGA by direct-sampling with a high-speed DAC (2.4 Giga-samples, 3rd nyquist zone), which significantly simplifies the RF hardware and allows generation of arbitrary waveforms. Control is performed by a Linux based Python TCP/IP server running on the Zynq's ARM processor that interfaces to MATLAB on the control PC.

\subsection{Bandwidth limitations of the NV-based electrometer}

The mechanism that determines the NV spin's sensitivity to high frequency electric fields at room temperature is limited by the spin-dependent readout rate. This rate is limited by the intersystem crossing process which is weakly temperature dependent due to its non-spin conserving property, and is $\approx$1/300~ns. Due to current experimental constraints such as limited photon collection efficiency and a limited bandwidth of the photodiode given the large dynamic range needed, the time constant on the LIA can then be set to match the bandwidth of the NV electrometer's spin readout of $\approx$3~MHz. Higher detection bandwidths can be achieved using single-shot spin readout at cryogenic temperatures.

\subsection{Excited-state optically detected magnetic resonance}
The spin physics of the \nv's excited state is identical with the \nv's ground state at temperatures above approximately 50~Kelvin~[25]. For purposes of sensing electric fields, the excited state is expected to be significantly less effective despite having 20~$\times$ greater transverse field sensitivity. This is attributed to shorter optical spontaneous lifetime (12~ns) and smaller ODMR contrast in the excited state. This analysis can be validated by substituting values from Fig.~7 into Eqn.~4.
\begin{figure} 
	\centering
		\includegraphics[width=90mm]{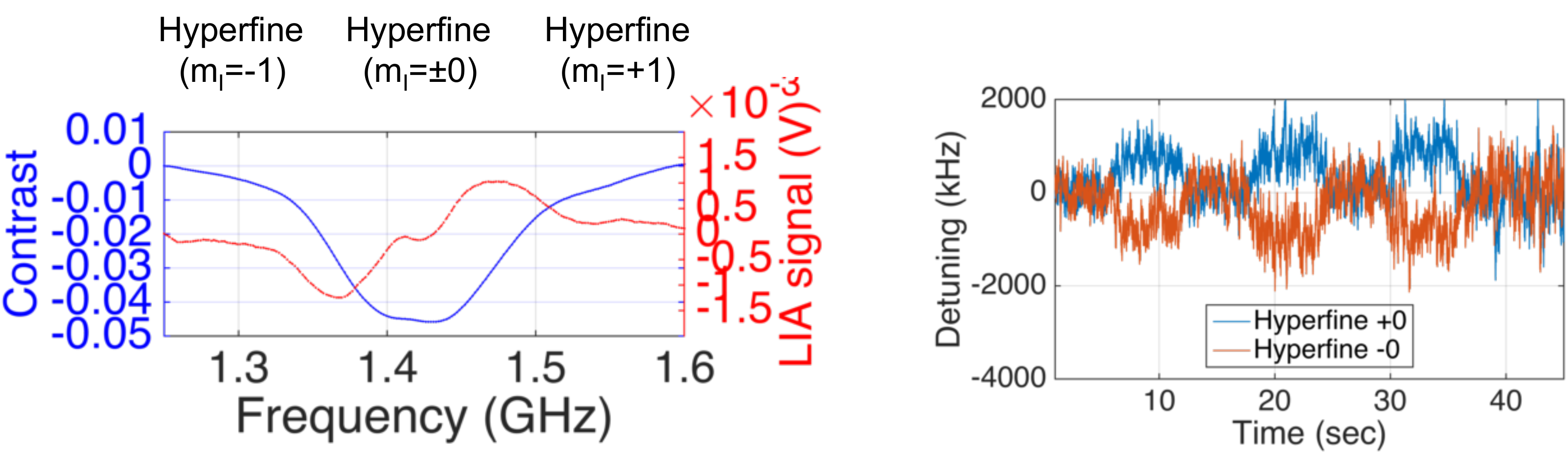}
	\caption{Excited-state ODMR spectra measured on an \nv~ensemble at zero magnetic field. Excited-state shifts due to pulsed electric fields applied to an \nv~ensemble at zero magnetic field. The sensitivity of this measurement approaches $300\frac{\text{V/cm}}{\sqrt{\text{Hz}}}$.
	}
\label{fig:efield_excited_shift}
\end{figure}

\begin{figure} 
	\centering
		\includegraphics[width=90mm]{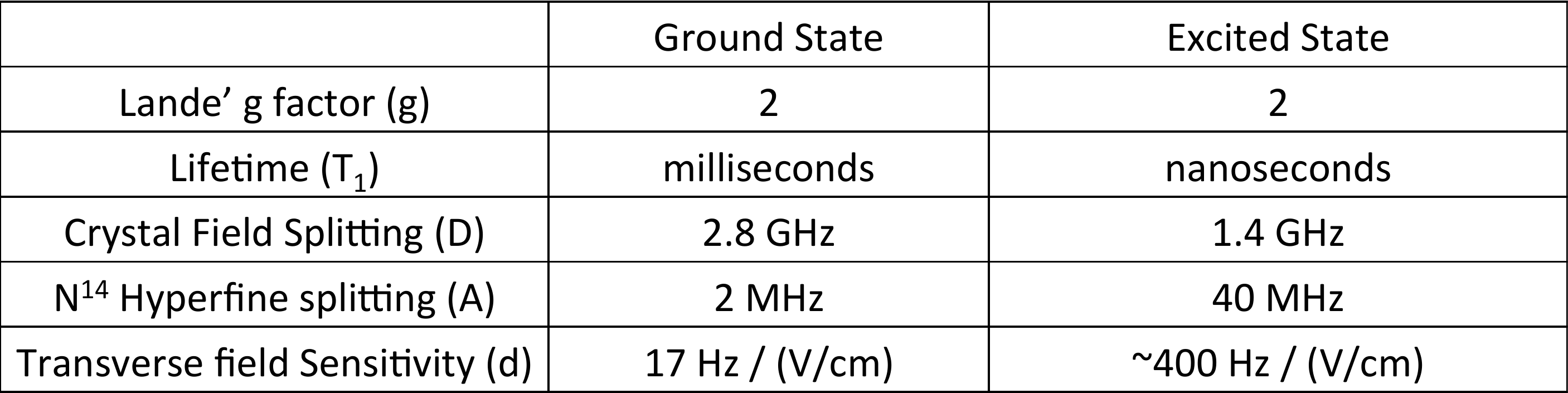}
	\caption{ Table outlining the major differences between the ground and excited state of the \nv~for electric field sensing.
	}
\label{fig:table}
\end{figure}

\end{document}